\documentclass[lettersize,journal]{IEEEtran}
\usepackage{amsmath,amsfonts}
\usepackage{array}
\usepackage{textcomp}
\usepackage{stfloats}
\usepackage{diagbox}
\usepackage{url}
\usepackage{verbatim}
\usepackage{graphicx}
\usepackage{adjustbox}
\usepackage{multirow}
\usepackage{booktabs}
\usepackage{makecell}
\usepackage{xcolor}
\usepackage{tikz}
\usepackage{pgf-pie,pgfplots,pgfplotstable}
\pgfplotsset{compat=1.18}
\usepackage{hyperref}
\usepackage{subcaption}
\usepackage{enumitem}
\usepackage{times}
\usepackage{latexsym}
\usepackage{microtype}
\usepackage{inconsolata}
\usepackage{algpseudocode}
\usepackage{colortbl}
\usepackage{siunitx}

\newcommand{\hhr}[1]{\textcolor{black}{#1}}

\makeatletter
\newcommand\footnoteref[1]{\protected@xdef\@thefnmark{\ref{#1}}\@footnotemark}
\makeatother

\def\BibTeX{{\rm B\kern-.05em{\sc i\kern-.025em b}\kern-.08em
    T\kern-.1667em\lower.7ex\hbox{E}\kern-.125emX}}
\usepackage{balance}

\usepackage{lipsum}
\usepackage[most]{tcolorbox}
\usetikzlibrary{positioning}
\tcbuselibrary{skins,breakable}
\newtcolorbox{mybox}[2][]{breakable,sharp corners, skin=enhancedmiddle jigsaw,parbox=false,
boxrule=0mm,leftrule=2mm,boxsep=0mm,arc=0mm,outer arc=0mm,attach title to upper,
after title={:\ }, coltitle=black,colback=gray!10,colframe=black, title={#2},
fonttitle=\bfseries,#1}

\begin{document}
 
% \title{Spontaneous Speech Generation with Extensive, Multilingual, and Diverse Data in the Wild}
\title{\hhr{\textsc{Emilia}}: A Large-Scale, Extensive, Multilingual, and Diverse Dataset for Speech Generation}
% \title{Spontaneous Speech Generation with Extensive, Multilingual, and Diverse Data in the Wild}

%names
\author{
Haorui~He$^{\star}$,
Zengqiang~Shang$^{\star}$,
Chaoren~Wang$^{\star}$,
Xuyuan~Li$^{\star}$,
Yicheng~Gu, \\
Hua~Hua,
Liwei~Liu, 
Chen~Yang, 
Jiaqi~Li,  
Peiyang~Shi, 
Yuancheng~Wang, \\
Kai~Chen, 
Pengyuan~Zhang$^{\dagger}$,
Zhizheng~Wu$^{\dagger}$,~\IEEEmembership{Senior Member,~IEEE}%
% affiliations
\thanks{Haorui He, Chaoren Wang, Yicheng Gu, Liwei Liu, Jiaqi Li, Yuancheng Wang, and Zhizheng Wu are with the Chinese University of Hong Kong, Shenzhen, China.}
\thanks{Zengqiang Shang, Xuyuan~Li, Hua~Hua, Chen~Yang, Peiyang~Shi, and Pengyuan~Zhang are with the Laboratory of Speech and Intelligent Information Processing, Institute of Acoustics, CAS, Beijing, China.}
\thanks{Xuyuan~Li, Hua~Hua, Chen~Yang, and Pengyuan~Zhang are also with the University of Chinese Academy of Sciences, Beijing, China.}
\thanks{Kai Chen is with Shanghai AI Laboratory, Shanghai, China.}
% roles
\thanks{$\star$: Equal contribution, and the names are listed in random order.}
\thanks{$\dagger$: Corresponding authors. Email: wuzhizheng@cuhk.edu.cn}
\thanks{This work is led by CUHK-Shenzhen and supported by the National Natural Science Foundation of China (No. 62376237), the 2023 Shenzhen Stability Science Program, the Shenzhen Science and Technology Program (No. ZDSYS20230626091302006), the Program for Guangdong Introducing Innovative and Entrepreneurial Teams (No. 2023ZT10X044), and the Postdoctoral Fellowship Program of China Postdoctoral Science Foundation (No. GZB20230811).}
}

\maketitle

\begin{abstract}
Recent advancements in speech generation have been driven by large-scale training datasets.
\hhr{However, current models struggle to capture the spontaneity and variability inherent in real-world human speech, as they are primarily trained on audio-book datasets limited to formal, read-aloud speaking styles.}
To address this limitation, we introduce \textit{Emilia-Pipe}, an open-source preprocessing pipeline designed to extract high-quality training data from valuable yet under-explored in-the-wild \hhr{sources} that capture spontaneous human speech in real-world contexts.
Using Emilia-Pipe, we construct \textit{Emilia}, which comprises over 101k hours of speech across six languages: English, Chinese, German, French, Japanese, and Korean. Furthermore, we expand Emilia to \textit{Emilia-Large}, a dataset exceeding 216k hours, making it \hhr{one of the largest open-source speech generation resources available.}
Extensive experiments show that Emilia-trained models produce markedly more spontaneous, human-like speech than those trained on traditional audio-book datasets, while matching their intelligibility. These models better capture diverse speaker timbres and the full spectrum of real-world conversational styles.
Our work also highlights the importance of scaling dataset size for advancing speech generation performance and validates the effectiveness of Emilia for both multilingual and crosslingual speech generation tasks.
\end{abstract}


\IEEEpeerreviewmaketitle

\section{Introduction}
\IEEEPARstart{I}{n} recent years, (zero-shot) speech generation research has witnessed significant advancements \cite{valle,ju2024naturalspeech,voicebox,borsos2023soundstorm,lajszczak2024base,seedtts}, with multiple models \hhr{driven by} large-scale training datasets. These advancements lead to improved voice quality, timbre similarity, and naturalness~\cite{amphion}. Nevertheless, the generated speech still falls short of replicating the spontaneity and variability characteristic of real-world human speech~\cite{ju2024naturalspeech,tan2021survey}.

A primary factor for this limitation is the reliance of current speech generation models on datasets derived from audio-books~\cite{librilight, mls}. Such datasets predominantly feature formal, read-aloud speaking styles, which contrast with the diverse and spontaneous nature of human speech in casual or conversational settings. Such real-world speech is characterized by a wide range of phenomena, including breathing sounds, pauses, stuttering, repetitions, variations in speaking rate and emotions. \hhr{\textit{Consequently, there remains a significant research gap for a dataset that encompasses a wider spectrum of speaking styles for speech generation model training.}}

However, directly utilizing in-the-wild speech data presents significant challenges due to variations in quality, such as frequent background noise or music, reverberation, overlapping speakers within a single sample, inconsistent speech lengths, and the absence of essential annotations like text transcriptions~\cite{AutoPrep,wenetspeech4tts}. 
Training speech generation models on such unprocessed raw data can result in degraded performance~\cite{AutoPrep,how_far_robust_vc,li2024sf}.
While previous studies~\cite{AutoPrep,wenetspeech4tts} have proposed automatic preprocessing pipelines to mitigate these issues, they heavily depend on proprietary models, which significantly limits their accessibility for the broader research community. Moreover, these pipelines are typically restricted to monolingual (i.e., Chinese-only) speech data, making them unsuitable for processing the multilingual speech data available in the wild. Finally, the computational efficiency of these approaches also remains undocumented, raising concerns about their practicality for building large-scale speech generation datasets. \textit{Therefore, there is an urgent need for an effective and open-source preprocessing pipeline that can efficiently handle multilingual in-the-wild speech data and enable large-scale dataset construction}.

% PIPE
In response, we present \textbf{\textit{Emilia-Pipe}}, the first open-source preprocessing pipeline specifically designed to leverage valuable yet underexplored in-the-wild multilingual speech data for constructing high-quality training datasets for spontaneous and human-like speech generation models. 
Emilia-Pipe comprises six core preprocessing steps: standardization, source separation, speaker diarization, fine-grained segmentation by voice activity detection (VAD), automated speech recognition (ASR), and filtering. 
In addition, Emilia-Pipe incorporates extensive engineering optimizations to enhance both robustness and efficiency. These features position Emilia-Pipe as an effective and efficient tool for building large-scale, multilingual speech datasets.

% DATA
Utilizing Emilia-Pipe, we introduce \textbf{\textit{Emilia}}, the first multilingual speech generation dataset constructed from in-the-wild speech data. Emilia comprises over 101k hours of 24 kHz speech across six languages: English (En), Chinese (Zh), German (De), French (Fr), Japanese (Ja), and Korean (Ko). To further scale up our dataset, we present \textbf{\textit{Emilia-Large}}, which expands the total duration to 216k hours.
Table~\ref{tab:speech_datasets} compares Emilia and Emilia-Large with several existing datasets. 

The main advantages of our proposed Emilia and Emilia-Large datasets are summarized below.
\begin{itemize}
\item \textbf{Extensive}: Emilia contains over 101k hours of speech data, and the Emilia-Large variant expands the size to 216k hours, which is \textbf{\textit{\hhr{one of the largest open-source speech generation datasets.}}}
\item \textbf{Multilingual}: The Emilia and Emilia-Large datasets cover six languages, supporting the training of \textbf{\textit{multilingual and crosslingual}} speech generation models. 
\item \textbf{Diverse}: Emilia and Emilia-Large distinguish themselves from prior datasets by centering on \textbf{\textit{spontaneous, in-the-wild speech}}. 
This ensures coverage of a broad spectrum of speaking styles, which is essential for training next-generation speech generation models capable of producing natural, spontaneous, and human-like speech.
\item \textbf{Dynamic}: The Emilia and Emilia-Large datasets uniquely feature an automatic and efficient processing pipeline, i.e., Emilia-Pipe, which \textbf{\textit{enables seamless expansion in both dataset size and language coverage}}, significantly accelerating dataset construction.
\end{itemize}

\begin{table*}[htbp]
    \centering
    \caption{A comparison of Emilia and Emilia-Large datasets with existing datasets for speech generation.}
    \label{tab:speech_datasets}
    \resizebox{\textwidth}{!}{
        \begin{tabular}{ccccccc}
            \toprule
            \textbf{Dataset} & \textbf{Data Source} & \textbf{Total Duration (hours)} & \textbf{Language} & \textbf{Samp. Rate (Hz)}  & \textbf{Dynamic} \\
            \midrule
            LJSpeech~\cite{ljspeech} & Audio-book & 24 & En & 22.05k  &   \\
            AutoPrepWild~\cite{AutoPrep} & In-the-wild & 39 & Zh & 24k/44.1k   & \checkmark (Proprietary)\\
            VCTK~\cite{vctk} & Studio Recording & 44 & En & 48k  &  \\
            AISHELL-3~\cite{aishell3}& Studio Recording  & 85 & Zh & 44.1k   &   \\
            LibriTTS~\cite{libritts} & Audio-book & 585 & En & 24k & &  \\
            GigaSpeech~\cite{gigaspeech}& In-the-wild & 10k & En & 16k &   \\
            \hhr{WenetSpeech~\cite{wenetspeech} }& In-the-wild & 10k & Zh & 16k &   \\
            WenetSpeech4TTS~\cite{wenetspeech4tts}& In-the-wild & 12k & Zh & 16k  & \checkmark (Proprietary) \\
            Libri-Heavy~\cite{libriheavy} & Audio-book & 50k & En & 16k  &  \\
            MLS~\cite{mls} & Audio-book & 51k & En/Fr/De/Nl/Es/It/Pt/Pl & 16k &  \\ 
            Libri-Light~\cite{librilight} & Audio-book & 60k & En & 16k  &  \\
            \hhr{SeamlessAlign~\cite{seamless}} & In-the-wild & 270k & 76 Languages & 16k  & \\
            \midrule
            Emilia & In-the-wild & 101k  & En/Zh/De/Fr/Ja/Ko& 24k   & \checkmark \\
            Emilia-Large & In-the-wild & 216k  & En/Zh/De/Fr/Ja/Ko& 24k   & \checkmark \\
            \bottomrule
        \end{tabular}
    }
\end{table*}

This work extends upon our previous research presented at IEEE Spoken Language Technology Workshop 2024~\cite{emilia}, introducing the following four key enhancements:
\begin{itemize}
    \item \textbf{Larger-scale Dataset}: We expanded the initial Emilia dataset to create Emilia-Large, a dataset more than twice the size of its predecessor.
    \item \textbf{Comparative Analysis of Audio-book and In-the-Wild Datasets}: We train identical models on traditional audio-book datasets and on the in-the-wild Emilia dataset, then compare their speech generation performance. \hhr{Results demonstrate that models trained with in-the-wild data achieve significantly higher speaker similarity and naturalness in the synthesized speech.}
    \item \textbf{Exploration of Data Scaling Laws in Speech Generation}: We conducted experiments to investigate the effect of dataset size on speech generation performance. The results highlight the importance of scaling dataset sizes.
    \item \textbf{Multilingual and Crosslingual Effectiveness Analysis}: We conduct experiments to validate the multilingual and crosslingual capabilities of Emilia, further demonstrating its applicability across six different languages.
\end{itemize}

Our code for Emilia-Pipe\footnote{\url{https://github.com/open-mmlab/Amphion/tree/main/preprocessors/Emilia}} as well as the Emilia and Emilia-Large datasets\footnote{\url{https://huggingface.co/datasets/amphion/Emilia-Dataset}} have been made publicly available to facilitate future research and ensure reproducibility. The remainder of this paper is organized as follows: 
Section~\ref{sec:related} reviews related work.
Section~\ref{sec:process_pipeline} describes the proposed Emilia-Pipe. 
Section~\ref{sec:dataset} details and analyzes our datasets.
Experimental results are presented in Section~\ref{sec:exp}. 
Section~\ref{sec:conclusion} concludes the paper.

% Additionally, during the course of this work, significant advancements in speech generation research have emerged using the initial Emilia dataset. These include, but are not limited to, MaskGCT~\cite{maskgct}, F5-TTS~\cite{f5tts}, and Vevo~\cite{vevo}, all of which provide open-source pre-trained models or code, serving as valuable contributions to the research community.

\section{Related Work}\label{sec:related}

\subsection{Speech Generation Datasets}
The size of datasets for speech generation has increased substantially over the years. 
Early datasets typically comprised tens of hours of speech. 
For example, LJSpeech \cite{ljspeech} contains 24 hours of speech data from a single speaker. 
The VCTK dataset \cite{vctk} includes 44 hours of speech data from 109 speakers, while AISHELL-3 \cite{aishell3} comprises approximately 85 hours of recordings from 218 speakers.
Subsequently, larger-scale datasets have emerged to enable research in zero-shot speech generation. 
For instance, LibriTTS \cite{libritts} contains 585.8 hours of speech data from audio-books, 
Later, the research community scaled up speech generation datasets to over 10k hours. 
Emerging datasets such as MLS \cite{mls} (51k hours), Libri-Light \cite{librilight} (60k hours), and Libri-Heavy \cite{libriheavy} (50k hours) significantly enhance the performance for zero-shot speech generation models. 
\hhr{However, these datasets are derived from audio-books and thus primarily capture formal, read-aloud speaking style, limiting their effectiveness for training natural and spontaneous speech generation models \cite{tan2021survey,tan2024naturalspeech}. While large-scale in-the-wild corpora such as GigaSpeech \cite{gigaspeech}, WenetSpeech \cite{wenetspeech}, and SeamlessAlign \cite{seamless} are available alternatives for broader speaking styles, their uncurated content and inconsistent speech quality can severely degrade generation performance if used without careful preprocessing \cite{AutoPrep,how_far_robust_vc}.}

To address this, two previous works \cite{wenetspeech4tts,AutoPrep} propose similar automated preprocessing pipelines for building speech generation datasets from in-the-wild data. However, as discussed earlier, these pipelines heavily rely on proprietary models and are restricted to Chinese-only speech data, with unknown computational efficiency.
To bridge this gap, we design and open-source an efficient pipeline, Emilia-Pipe, which can rapidly process large-scale raw multilingual speech data to facilitate large-scale dataset construction.

\subsection{Speech Generation Models}

\hhr{Traditional speech generation models, such as Tacotron \cite{wang2017tacotron}, FastSpeech \cite{ren2019fastspeech}, are limited by smaller datasets with only tens of hours of speech data. 
Recent advancements in speech generation, driven by large-scale speech datasets, have significantly improved voice quality, timbre similarity, and naturalness in zero-shot speech generation using only a short reference speech sample of a few seconds. For example, VALLE \cite{valle}, VoiceBox \cite{voicebox}, and SoundStorm \cite{borsos2023soundstorm} use more than 50k hours of speech data for training. Subsequent models, such as NaturalSpeech3 \cite{ju2024naturalspeech}, BaseTTS \cite{lajszczak2024base}, and Seed-TTS \cite{seedtts}, further expanded dataset sizes to over 100k hours. 
Notably, BaseTTS reported the ``emergent abilities'' of TTS models: as dataset sizes scaled, the models can render complex prosody patterns such as emotions based on textual cues without explicit labels.
In this work, we introduce Emilia-Large, one of the largest open-source datasets for speech generation of 216k hours, to further advance the development of speech generation.}

\begin{figure*}[t]
    \centering
    \includegraphics[width=\linewidth]{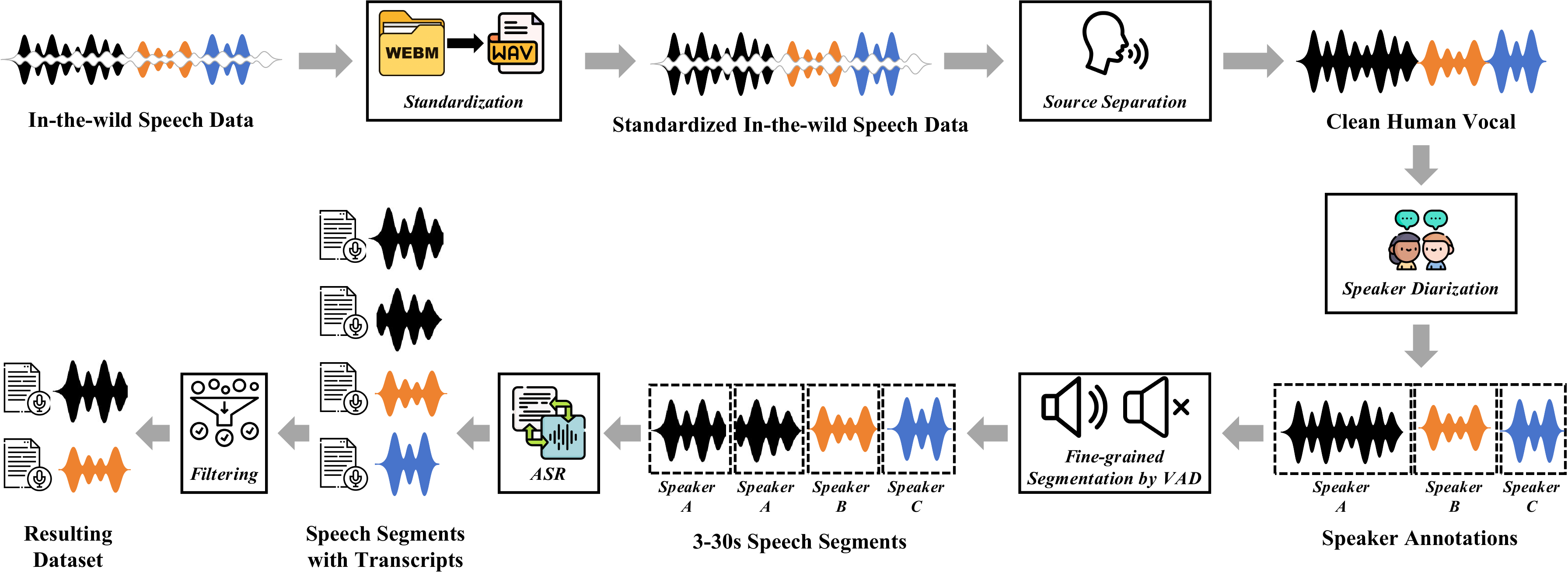}
    \caption{An overview of the Emilia-Pipe pipeline. It consists of six steps, namely, standardization, source separation, speaker diarization, fine-grained segmentation by voice activity detection (VAD), automated speech recognition (ASR), and filtering.}
    \label{fig:process_pipeline}
\end{figure*}

\section{The Emilia-Pipe Processing Pipeline} \label{sec:process_pipeline}

As discussed above, source speech data in-the-wild need to be processed to be leveraged for training speech generation models. Therefore, we design an automatic preprocessing pipeline, Emilia-Pipe, for transforming in-the-wild multilingual speech data into high-quality training datasets. As illustrated in Fig.~\ref{fig:process_pipeline}, Emilia-Pipe includes six steps: standardization, source separation, speaker diarization, fine-grained segmentation by voice activity detection (VAD), automated speech recognition (ASR), and filtering.
This section details the six steps of Emilia-Pipe and evaluates its performance.

\paragraph{Standardization}
The source speech data in-the-wild vary in encoding formats, sampling rates, and other characteristics. To standardize the collected data, we convert all samples to WAV files, set them to a mono channel, and resample to 24~kHz. We set the sample width to 16 bits and adjust the target decibels relative to full scale to -20~dBFS. The actual gain is constrained within -3 to 3~dB to ensure appropriate volume without distortion. Finally, we normalize the waveform by dividing each sample by the maximum amplitude, ensuring values range between -1 and 1. This step ensures a consistent data format for further processing.

\paragraph{Source Separation} The source speech data in-the-wild often contain background music and noise, which negatively impact the performance of speech generation models~\cite{voicebox, how_far_robust_vc}. To address this issue, we employ source separation techniques to extract clean human vocals. Specifically, we utilize the open-source pre-trained Ultimate Vocal Remover model \cite{mdxnet}, UVR-MDX-Net Inst 3\footnote{\url{https://github.com/TRvlvr/model_repo/releases/tag/all_public_uvr_models}}. This model achieves a high signal-to-distortion ratio (SDR) of 11.15 for vocal separation on the Synth MVSep dataset \cite{mvsep}. Using this model, we effectively separate human vocals for further processing.

\paragraph{Speaker Diarization} After extracting clean human vocals from the source speech data, we apply speaker diarization techniques to partition long-form speech data into multiple utterances based on the speaker. This process generates a series of speech segments, with each segment ideally containing only one speaker, ensuring compatibility with existing datasets for speech generation. To achieve this, we leverage the PyAnnote speaker diarization 3.1 pipeline.\footnote{\url{https://github.com/pyannote/pyannote-audio}}, which includes three core components: speaker segmentation, speaker embedding, and clustering, achieving state-of-the-art (SOTA) speaker diarization performance~\cite{pyannote2.1, pyannote_model}. The output is a list of temporal annotations indicating the start and end times of the single-speaker segments.

\paragraph{Fine-grained Segmentation by VAD}\label{sec:vad}
Although the speaker diarization pipeline provides a coarse segmentation for the source speech data, the resulting utterances may still be too long for training speech generation models. To address this, we use a VAD model to further segment the utterances into smaller segments ranging from 3 to 30 seconds. Specifically, we concatenate consecutive chunks containing voice activity from the same speaker. We leverage the open-source library Silero-VAD\footnote{\url{https://github.com/snakers4/silero-vad}}, which achieves a ROC-AUC score of 0.99 on the LibriParty dataset.\footnote{\url{https://github.com/speechbrain/speechbrain/tree/develop/recipes/LibriParty/generate_dataset}}

\paragraph{ASR}\label{sec:asr} The absence of text transcriptions impedes the direct use of in-the-wild dataset for TTS. Therefore, we use ASR techniques to transcribe the segmented speech data. Considering the trade-off among speed, robustness, and accuracy, we employ the medium version of the Whisper model~\cite{whisper}, a SOTA multilingual ASR model capable of robust speech translation and language identification. To further enhance efficiency, we leverage WhisperX \cite{whisperx}, which builds on the faster-whisper backend\footnote{\url{https://github.com/SYSTRAN/faster-whisper}} and the CTranslate2 engine.\footnote{\url{https://github.com/OpenNMT/CTranslate2}}
This setup is up to four times faster than the official Whisper implementation while maintaining comparable accuracy and using less memory. Additionally, we omit the original model's inherent VAD component by using the outputs in the last step to avoid redundant processing and develop batched inference to transcribe the speech data in parallel. These improvements allow our ASR step to achieve accurate text transcriptions for the speech data with high efficiency.

\begin{table*}[t!]
    \centering
    \caption{\hhr{Processing results of Emilia-Pipe.}}
    \label{tab:experimental_results}
    \resizebox{\textwidth}{!}{
    \begin{tabular}{lcccccccc}
        \toprule
        \multirow{2}{*}{\textbf{Processing Steps}} & \multicolumn{3}{c}{\textbf{Duration (s)}} & \multicolumn{3}{c}{\hhr{\textbf{DNSMOS $\uparrow$}}} & \multirow{2}{*}{\textbf{Clips}} & \multirow{2}{*}{\textbf{Total Duration (hours)}}\\
        \cmidrule(lr){2-4} \cmidrule(lr){5-7}
         & \textbf{min} & \textbf{max} & \textbf{avg \textpm~std} & \textbf{min} & \textbf{max} & \textbf{avg \textpm~std} & & \\
        \midrule
        Source Speech & 20.09 & 3596.27 & 547.80 \textpm~653.92 & 1.04 & 3.51 & 2.52 \textpm~0.68 & 4383 & 666.94 (100.00\%)\\  
        \midrule
        + Source Separation & 20.09 & 3596.27 & 547.80 \textpm~653.92 & 0.96 & 3.54 & 2.88 \textpm~0.50 & 4383 & 666.94 (100.00\%)\\ 
        + Speaker Diarization  & 0.02 & 1955.93 & 4.30 \textpm~10.73 & 0.64 & 3.69 & 2.50 \textpm~0.69 & 431876 & 515.50 (77.29\%)\\  
        + Fine-grained Segmentation by VAD & 3.00 & 30.00 & 9.05 \textpm~5.25 & 0.80 & 3.69 & 2.92 \textpm~0.50 & 207709 & 522.03 (78.27\%)\\
        + ASR & 3.00 & 30.00 & 9.16 \textpm~5.25 & 0.97 & 3.68 & 2.98 \textpm~0.44 & 172923 & 439.80 (65.94\%)\\
        + Filtering & 3.00 & 30.00 & 9.68 \textpm~5.17 & 3.00 & 3.68 & 3.26 \textpm~0.14 & 96117 & 258.44 (38.75\%) \\
        \midrule
        \multicolumn{9}{l}{\textbf{Total Processing Time}: 240.5 mins; \textbf{Real-Time Factor (RTF)}: 0.006} \\
        \bottomrule
    \end{tabular}
    }
\end{table*}

\paragraph{Filtering}
Real-world noises may not be completely handled by source separation, the ASR step may introduce errors, and some source speech data may be of low quality~\cite{AutoPrep}. Therefore, to ensure the quality of the resulting dataset, we apply the following filtering criteria.\footnote{Please note that the filtering criteria can be adjusted to fit the specific needs of different use cases.} Firstly, we utilize the language identification results from the Whisper model in the ASR step. We discard speech data that are not predicted to belong to our target languages (En, Zh, De, Fr, Ja, Ko) or have a language identification confidence lower than 80\%.
Secondly, we use the DNSMOS P.835 OVRL score~\cite{dnsmos835} (hereafter referred to as DNSMOS score for brevity) to estimate the overall speech quality. 
This non-intrusive metric reflects the overall quality of the speech data and is highly correlated with human ratings~\cite{dnsmos835}.
\hhr{Following ITU-T P.835 \cite{ITU-T}, we retain only utterances whose DNSMOS $\geq$ 3.0, the lowest ``fair'' quality anchor, thereby guaranteeing that every training sample meets a minimum bar accepted for high-fidelity speech synthesis.}
Finally, for each source speech data, we compute the average character duration over its corresponding segments. We consider segments with an average phone duration outside 1.5 times the interquartile range (IQR) above the third quartile or below the first quartile as outliers and discard the speech data for these segments.
After filtering, we obtain the resulting dataset.

\paragraph{Performance Evaluation} \label{sec:pip_exp}
\hhr{To evaluate the effectiveness of Emilia-Pipe, we processed 666.94 hours of randomly selected source speech data through the pipeline. Table~\ref{tab:experimental_results} illustrates the impact of each processing step within Emilia-Pipe.}

\hhr{Before processing, the raw source speech data is highly variable, with clip durations ranging from 20.09 to 3596.27 seconds (averaging 547.80 seconds $\pm$ 653.92 seconds), and DNSMOS scores spanning 1.04 to 3.51 (averaging 2.52 $\pm$ 0.68), reflecting varied speech quality in-the-wild.
Applying source separation preserves the total duration but improves the average DNSMOS score to 2.88 $\pm$ 0.50. Speaker diarization segments multi-speaker audio, increasing the number of clips to 431,876 while reducing the total duration to 515.50 hours (77.29\% of the original). Subsequent VAD steps further refine the dataset, constraining all segment durations to between 3 and 30 seconds. The final filtering step yields 96,117 high-quality clips totaling 258.44 hours (38.75\% of the original data), with a notably improved DNSMOS score of 3.26 $\pm$ 0.14, indicating low variability and superior speech quality. These results demonstrate Emilia-Pipe’s capability to convert raw speech data in the wild into a high-quality dataset for speech generation.}

\hhr{This experiment was conducted on a server equipped with eight NVIDIA RTX 4090 GPUs, running eight independent processes.\footnote{Processing speed may vary depending on hardware and data characteristics; reported figures are provided for reference.} On this server, processing the 666.94 hours of source speech data took 240.5 minutes, resulting in a Real-Time Factor (RTF) of approximately 0.006 (i.e., 1 second of source speech data processed in 0.006 seconds). This high efficiency underscores Emilia-Pipe's suitability for large-scale speech dataset preprocessing, enabling fast training data preparation for speech generation models.}

\section{The Emilia and Emilia-Large Dataset}\label{sec:dataset}
Leveraging Emilia-Pipe, we are able to construct speech generation datasets derived from in-the-wild speech data. In this section, we describe our constructed Emilia dataset and the extended Emilia-Large dataset. These datasets contain in-the-wild speech data in six languages (En, Zh, De, Fr, Ja, Ko), processed by Emilia-Pipe. Duration statistics for each language in the datasets are provided in Fig.~\ref{fig:dataset_stats}.

\begin{figure*}[t]
    \centering
    \begin{subfigure}[t]{0.84\columnwidth}
        \resizebox{\textwidth}{!}{
            \begin{tikzpicture}
                \definecolor{color1}{HTML}{447cac}
                \definecolor{color2}{HTML}{88ce9b}
                \definecolor{color3}{HTML}{e3f79b}
                \definecolor{color4}{HTML}{fae28c}
                \definecolor{color5}{HTML}{f1874b}
                \definecolor{color6}{HTML}{c42d40}
                \pie[
                    text=legend,
                    radius=3,
                    color={color1, color2, color3, color4, color5, color6},
                    explode=0.1, % Adds a small offset to the slices
                    sum=auto, % Automatically sums up to 100%
                    before number=\phantom{0}, % Adds a leading zero for consistency
                    after number=\% % Appends a percentage sign to the numbers
                ]{
                    46.77/En: 46.8k, 
                    49.87/Zh: 49.9k, 
                    1.59/De: 1.6k, 
                    1.38/Fr: 1.4k, 
                    1.72/Ja: 1.7k, 
                    0.22/Ko: 0.2k
                }
            \end{tikzpicture}
        }
        \caption{Emilia}
        \label{fig:dataset_stats_a}
    \end{subfigure}
    % \vspace{10pt} % 调整两个图之间的垂直间距
    \begin{subfigure}[t]{0.93\columnwidth}
        \resizebox{\textwidth}{!}{
            \begin{tikzpicture}
                % Define colors using hex codes
                \definecolor{color1}{HTML}{447cac}
                \definecolor{color2}{HTML}{88ce9b}
                \definecolor{color3}{HTML}{e3f79b}
                \definecolor{color4}{HTML}{fae28c}
                \definecolor{color5}{HTML}{f1874b}
                \definecolor{color6}{HTML}{c42d40}
                \pie[
                    text=legend,
                    radius=3,
                    color={color1, color2, color3, color4, color5, color6},
                    explode=0.1, % Adds a small offset to the slices
                    sum=auto, % Automatically sums up to 100%
                    before number=\phantom{0}, % Adds a leading zero for consistency
                    after number=\% % Appends a percentage sign to the numbers
                ]{
                    61.95/En: 134.1k (\textbf{2.9x}), 
                    26.49/Zh: 57.3k (\textbf{1.1x}), 
                    3.15/De: 6.8k (\textbf{4.3x}), 
                    3.80/Fr: 8.2k (\textbf{6.0x}), 
                    1.17/Ja: 2.5k (\textbf{1.5x}), 
                    3.44/Ko: 7.4k (\textbf{34.3x})
                }
            \end{tikzpicture}
        }
        \caption{Emilia-Large}
        \label{fig:dataset_stats_b}
    \end{subfigure}
    
    \caption{\hhr{Duration statistics (in hours) of the speech data in Emilia and Emilia-Large by language. The numbers in parentheses indicate the scaling factor (multiples) of the speech data in Emilia-Large compared to the original Emilia dataset.}}
    \label{fig:dataset_stats}
\end{figure*}

\subsection{The Emilia Dataset}

\subsubsection{Overview} Using Emilia-Pipe, we construct Emilia from in-the-wild speech data sourced from a vast collection of video and podcast platforms on the Internet. This data covers various speaking styles such as \hhr{audio-books, drama, interviews, talk shows, and commentary}, thereby capturing a wide array of real human speaking styles. After processing, Emilia includes a total of 101,654 hours of multilingual speech data across six different languages. 
 
\subsubsection{Dataset Analysis}
To validate the quality and diversity of Emilia, we conduct respective analyses.

\paragraph{Quality}
To evaluate the quality, we compared Emilia with several existing datasets using DNSMOS scores. Table~\ref{tab:dnsmos_results} presents the speech quality comparison between Emilia and several existing datasets. Emilia achieves a DNSMOS score of 3.26, ranking third among all datasets. The results indicate that, despite being sourced from source speech data in-the-wild, after preprocessing, the speech quality of Emilia is comparable to existing datasets sourced from studio recordings or audio-books and outperforms the existing datasets sourced from unprocessed in-the-wild speech data.

\begin{table}[t]
    \centering
    \caption{DNSMOS Scores of Emilia and ten existing datasets. The scores for LJSpeech, AutoPrepWild, AISHELL-3, LibriTTS, and WenetSpeech, are derived from~\cite{AutoPrep}. The score for Libri-Light is computed from its official ``small'' subset, and the score for WenetSpeech4TTS is computed from its official ``basic'' subset. The scores for MLS and Emilia are computed from a randomly sampled 600-hour subset.}
    \label{tab:dnsmos_results}
    \begin{tabular}{cc}
        \toprule
        \textbf{Dataset} & \hhr{\textbf{DNSMOS $\uparrow$}} \\
        \midrule
        LJSpeech~\cite{ljspeech} & 3.30 \textpm~0.17 \\
        AutoPrepWild~\cite{AutoPrep} & 3.24 \textpm~0.21 \\
        VCTK~\cite{vctk} & 3.20 \textpm~0.18 \\
        AISHELL-3~\cite{aishell3} & 3.15 \textpm~0.17 \\
        LibriTTS~\cite{libritts} & 3.25 \textpm~0.19 \\
        GigaSpeech~\cite{gigaspeech} & 2.52 \textpm~0.54 \\
        \hhr{WenetSpeech~\cite{wenetspeech}} & 2.43 \textpm~0.55 \\
        WenetSpeech4TTS~\cite{wenetspeech4tts} & 3.18 \textpm~0.22 \\
        MLS~\cite{mls} & \textbf{3.33 \textpm~0.19} \\
        Libri-Light~\cite{librilight} & 3.25 \textpm~0.26 \\
        \midrule
        Emilia & 3.26 \textpm~0.14 \\
        \bottomrule
    \end{tabular}
\end{table}

\begin{table}[thbp]
\centering
\caption{Distribution of speaking style domains in Emilia.}
\label{tab:emilia_categories}
\hhr{
\begin{tabular}{cc}
\toprule
Domain & Percentage \\
\midrule
Audio-book & 9.40\% \\
Commentary & 17.75\% \\
Documentary & 11.46\% \\
Drama & 12.09\% \\
Interview & 0.93\% \\
News & 2.77\% \\
Reading & 15.60\% \\
Talk & 13.02\% \\
Variety & 6.19\% \\
Others & 10.79\% \\
\bottomrule
\end{tabular}
}
\end{table}

\begin{figure}[t]
    \centering
    \begin{subfigure}[t]{0.75\columnwidth}
        \includegraphics[width=\textwidth]{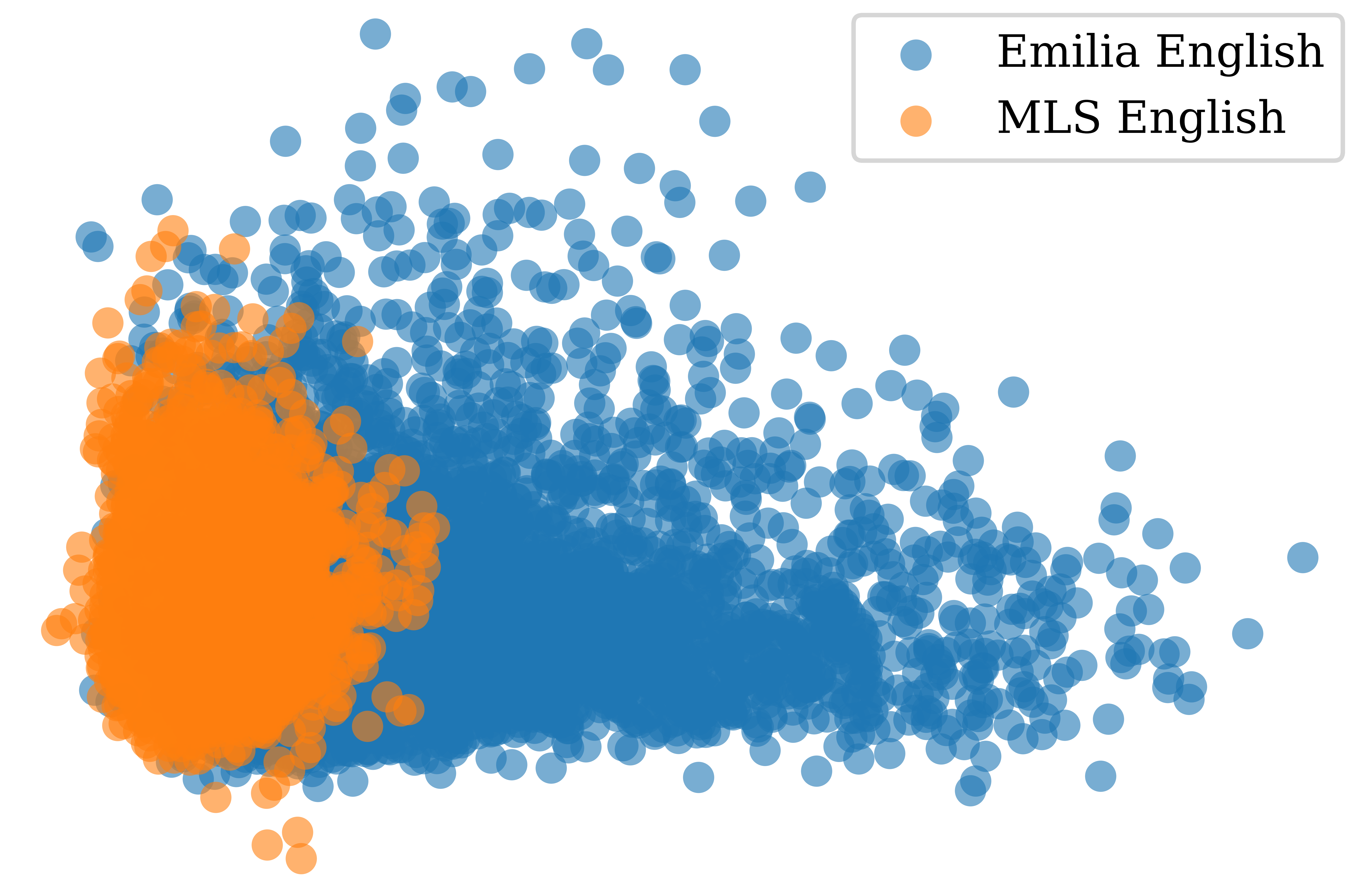}
        \caption{Acoustic diversity}
        \label{fig:acoustic}
    \end{subfigure}
    \begin{subfigure}[t]{0.75\columnwidth}
        \includegraphics[width=\textwidth]{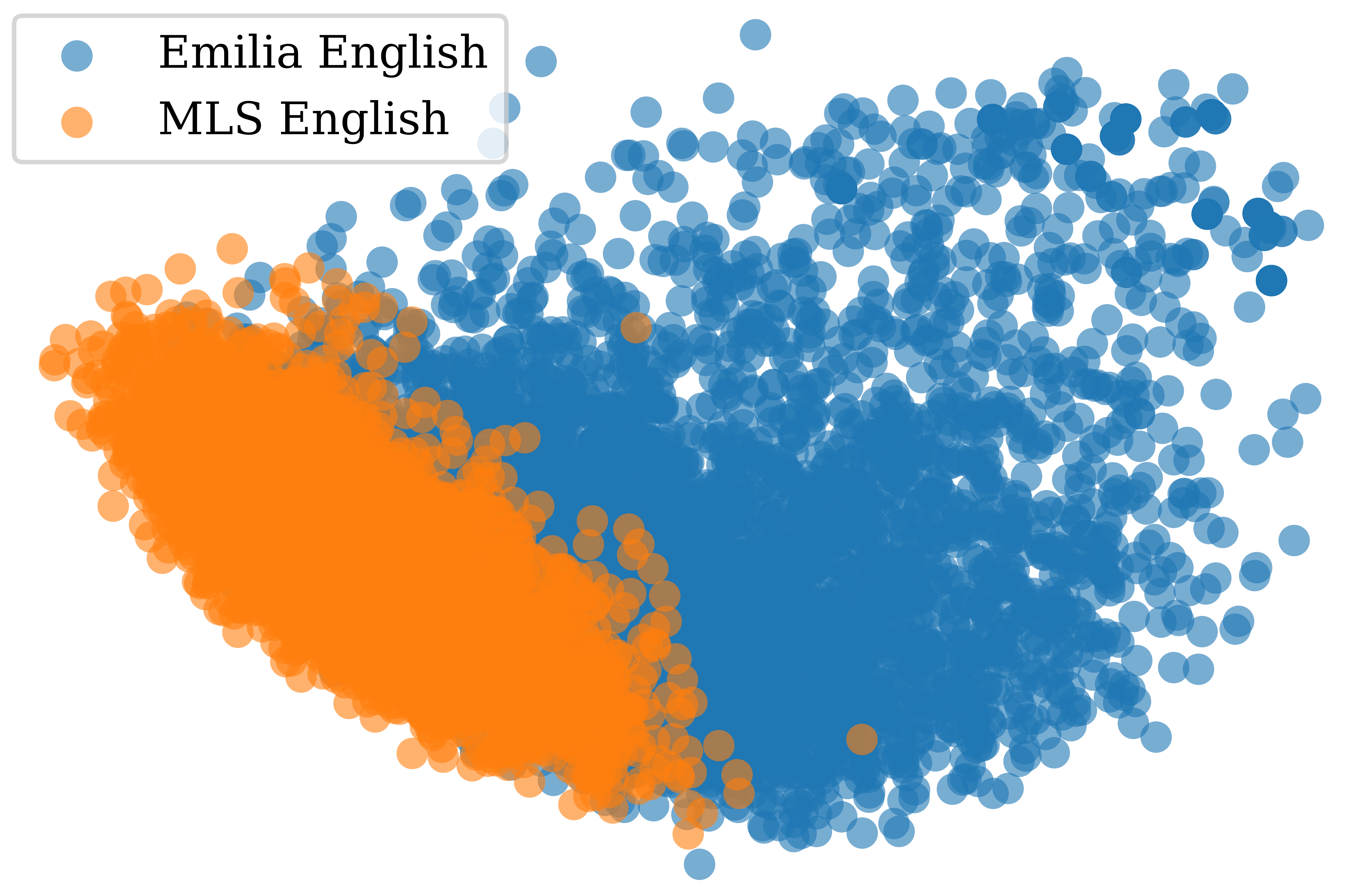}
        \caption{Semantic diversity}
        \label{fig:semantic}
    \end{subfigure}
    
    \caption{A comparison of acoustic and semantic diversities between Emilia and MLS datasets.}
    \label{fig:combined}
\end{figure}

\paragraph{Diversity}
% Emilia comprises a collection of speech data from a wide range of video and podcast platforms. 
\hhr{
To quantify the diversity of speaking style domains within Emilia, we follow WenetSpeech \cite{wenetspeech} by classifying the source speech data into ten domains: Audio-book, Commentary, Documentary, Drama, Interview, News, Reading, Talk, Variety, and Others.}

\hhr{This classification is performed by analyzing textual information extracted from source URLs and metadata (such as hashtags) using Gemma-3-12B-it.\footnote{\url{https://huggingface.co/google/gemma-3-12b-it}} As shown in Table~\ref{tab:emilia_categories}, Emilia demonstrates a well-balanced distribution of speaking styles across most domains}

\hhr{
To illustrate the diversity advantage of Emilia over traditional audio-book datasets, we perform a comparative analysis between Emilia and the MLS dataset in both acoustic and semantic feature spaces.
Specifically, we randomly select 5,000 samples each from the English subsets of Emilia and MLS, denoted as Emilia English and MLS English.}

For acoustic features, we leverage a pre-trained WavLM model \cite{chen2022wavlm} to extract 768-dimensional embeddings for each speech sample, which captures various acoustic attributes such as speaker identity, emotion, and prosody. 
These high-dimensional features are then reduced to two dimensions using Principal Component Analysis (PCA) \cite{pca}. 
As shown in Fig.~\ref{fig:combined}(a), Emilia displays a wider distribution compared to the more compact clustering of MLS, highlighting Emilia's richer coverage in acoustic characteristics.

For semantic diversity, we employ a pre-trained Sentence-BERT model \cite{sentence_bert} to generate 768-dimensional text embeddings from the transcripts of each sample. These embeddings capture the semantic content of these samples. Similar to above, we reduce the dimensions of these embeddings to two using PCA. Fig.~\ref{fig:combined}(b) demonstrates that Emilia spans a broader semantic coverage than MLS.

\subsection{The Emilia-Large Dataset}
Given the significant advancements in large-scale speech generation~\cite{emilia}, we are motivated to further scale up the volume of the training dataset for speech generation models to investigate the data scaling laws in speech generation and potentially enhance model performance. In this work, we expand the initial Emilia dataset to an even larger scale, introducing Emilia-Large, which contains a total of 216,313 hours of spontaneous and human-like speech data. Emilia-Large builds upon the same construction methodology as Emilia using our proposed Emilia-Pipe but doubles the total dataset size.

The expansion from Emilia to Emilia-Large primarily uses YODAS2~\cite{li2023yodas} as the data source, termed Emilia-YODAS. 
The relationship between Emilia and Emilia-Large is illustrated in Fig.~\ref{fig:venn_emilia_large}. 
YODAS2 is a large-scale (500k hours) real-world speech collection from YouTube videos with CC-BY-3.0 licenses in more than 100 languages. We selectively downloaded and processed data from YODAS2 and other smaller sources in the six languages of the original Emilia dataset.\footnote{Unused data of other languages in YODAS2 can also be seamlessly processed by Emilia-Pipe for speech generation training. Their effectiveness may depend on the ASR model's performance in each language.}
We process the source speech data using our proposed Emilia-Pipe, with only one alteration: changing the DNSMOS filtering threshold for the incremental data to 2.4 to align with~\cite{AutoPrep} and preserve more data. We make the Emilia-YODAS publicly available under the CC-BY-4.0 license.

Fig.~\ref{fig:dataset_stats}(b) demonstrates the duration statistics for each language in Emilia-Large. It is observed that the key distinction of Emilia-Large compared to Emilia is its significantly improved inclusion of low-resource languages, especially for German (De), French (Fr), and Korean (Ko). Specifically, compared to the original Emilia dataset, the data for these three languages has been expanded several-fold (4.3 times for De, 6.0 times for Fr, and 34.3 times for Ko, respectively). This enhancement addresses the relatively limited data volume of these languages in Emilia, improving support for multilingual and crosslingual speech generation tasks.

\begin{figure}[t]
\centering
\resizebox{0.62\columnwidth}{!}{
\begin{tikzpicture}
\definecolor{color1}{HTML}{E0F7FA} % Pale Blue
\definecolor{color3}{HTML}{B0E0E6} % Pale Green
\definecolor{color2}{HTML}{D1E8E2}
    % Draw the larger ellipse for Emilia-Large
    \draw[fill=color1] (0,0) ellipse (4cm and 3cm);
    
    % Draw the smaller circles inside for Emilia and Emilia-YODAS
    \draw[fill=color2] (-2,0) circle (2cm);
    \draw[fill=color3] (2,0) circle (2cm);
    
    % Add labels
    \node at (0, 2.25) {\parbox{4cm}{\centering \textbf{Emilia-Large} \\ (216k hours)}};
    
    % Annotate the parts with arrows for better clarity
    \node at (2, 0) [align=center] {Emilia-YODAS \\ (CC-BY-4.0, \\ 114k hours)};
    \node at (-2, 0) [align=center] {Emilia \\ (CC-BY-NC-4.0, \\ 101k hours)};    
    % Adding additional text for context
    \node at (0, -2.25) [align=center] {Other sources};
\end{tikzpicture}
}

\caption{The relationship between Emilia and Emilia-Large. Emilia-Large is an extended version of Emilia, incorporating additional data primarily processed from YODAS2.}
\label{fig:venn_emilia_large}
\end{figure}

\section{Experiments}\label{sec:exp}
In this section, we conduct experiments to address the following evaluation questions (EQs) to validate the strengths of Emilia in terms of diversity (EQ1), extensiveness (EQ2), and multilingual utility (EQ3):

% First, how does training speech generation models on Emilia, an in-the-wild dataset, compare in performance how to training on existing audio-book datasets (EQ1)? Second, what are the data scaling laws in speech generation, that is, how does dataset size affect model performance when the number of parameters is fixed (EQ2)? Third, how effective is Emilia for training multilingual and crosslingual speech generation models (EQ3)?

\begin{itemize}
    \item \textbf{EQ1:} \hhr{Is Emilia more effective than existing audio-book datasets for training spontaneous and human-like speech generation model?}
    \item \textbf{EQ2:} What are the data scaling laws in speech generation, i.e., the impact of dataset size on speech generation performance with a fixed number of model parameters?
    \item \textbf{EQ3:} How effective is Emilia for training multilingual and crosslingual speech generation models?
\end{itemize}

\subsection{Comparison of Audio-book and In-the-Wild Datasets (EQ1)} \label{sec:eq1}
\hhr{To address EQ1, we compare the performance of TTS models trained on existing audio-book datasets to the performance of models trained on Emilia.}

\paragraph{Baselines}
We implement two SOTA TTS models as baselines: 
(1) AR+SoundStorm \cite{maskgct}: 
A two-stage model where a LLaMA-style autoregressive (AR) generative transformer first predicts semantic tokens using text and prompt semantic tokens as input. Then a SoundStorm-based semantic-to-acoustic model \cite{borsos2023soundstorm} that generates acoustic tokens conditioned on the predicted semantic tokens.
(2) VoiceBox \cite{voicebox}: A non-autoregressive (NAR) speech generation model that leverages flow matching to predict mel-spectrograms.
For comprehensive details on these models, we refer readers to their respective publications.

% Training Sets
\paragraph{Training Sets} We evaluate the performance of TTS models trained on two English datasets: the English subset of Emilia (hereafter referred to as Emilia-En for brevity) and the MLS dataset, a high-quality corpus derived from audio-books. The Emilia-En dataset consists of approximately 46k hours of English speech, while the MLS dataset contains 45k hours, rendering their sizes comparable.

% Evaluation Sets
\paragraph{Evaluation Sets}
To ensure a comprehensive evaluation, we employ two evaluation sets in two domains: 
(1) LibriSpeech-Test: This evaluation set includes 1,200 speech samples in formal reading styles similar to those in the MLS dataset.
(2) Emilia-Test: This evaluation set consists of 600 speech samples in spontaneous, human-like speaking styles akin to those in Emilia.
Both evaluation sets are unseen by the baseline models during training.

% Metrics
\paragraph{Metrics}
We conduct both objective and subjective evaluations to assess the performance of the baseline models.

\hhr{For objective evaluation, we focus on the following aspects.
(1) Intelligibility: This is measured by the Word Error Rate (WER) between the generated speech transcription and the target text. For LibriSpeech-Test, we use a fine-tuned HuBERT-Large ASR model\footnote{\url{https://huggingface.co/facebook/hubert-large-ls960-ft}} to transcribe the generated speech. For Emilia-Test, we use the Whisper-Medium model \cite{whisper}, which offers greater robustness for in-the-wild speech.
(2) Similarity: This is measured by Speaker Similarity Score (S-SIM) between the generated speech and the reference speech using the WavLM-TDCNN speaker embedding model.\footnote{\url{https://github.com/microsoft/UniSpeech/tree/main/downstreams/speaker_verification}}
(3) Naturalness: This is measured using the Fréchet Speech Distance (FSD), which quantifies the distance between the distributions of generated and real speech samples in a feature space \cite{voicebox}. We employ the SOTA emotion representation model, emotion2vec \cite{emotion2vec}, to compute FSD and evaluate the emotional naturalness of the generated speech.}

For the subjective evaluation, we randomly select sixteen samples, eight from the LibriSpeech-Test and eight from the Emilia-Test evaluation set. Twelve proficient English speakers serve as evaluators. The subjective evaluation includes:
(1) Speaker Similarity: We employ the Similarity Mean Opinion Score (SMOS) to assess the speaker similarity of the generated speech to the reference speech. The SMOS scale ranges from 1 to 5, with steps of 0.5.
(2) Comparative Naturalness: We use the Comparative Mean Opinion Score (CMOS) to evaluate the comparative naturalness of the generated speech against the reference speech. The CMOS scale ranges from -3 (indicating the generated speech is much worse than the reference speech) to 3 (indicating the generated speech is much better than the reference speech), with steps of 1.

\begin{table*}[t]
\centering
\caption{Objective and subjective evaluation results of TTS models trained on Emilia-En and MLS on LibriSpeech-Test and Emilia-Test evaluation sets. The best results for each model are highlighted in bold.}
\label{tab:combined_evaluation}
\begin{tabular}{ccccccccccccc}
\toprule
\multirow{2}{*}{\textbf{Model}} & \multirow{2}{*}{\textbf{Training Set}} & \multicolumn{5}{c}{\textbf{LibriSpeech-Test}} & \multicolumn{5}{c}{\textbf{Emilia-Test}} \\ 
\cmidrule(lr){3-7} \cmidrule(lr){8-12}
 &  & \hhr{\textbf{WER $\downarrow$}} & \hhr{\textbf{S-SIM $\uparrow$}} & \hhr{\textbf{FSD $\downarrow$}} & \hhr{\textbf{CMOS $\uparrow$}} & \hhr{\textbf{SMOS $\uparrow$}} & \hhr{\textbf{WER $\downarrow$}} & \hhr{\textbf{S-SIM $\uparrow$}} & \hhr{\textbf{FSD $\downarrow$}} & \hhr{\textbf{CMOS $\uparrow$}} & \hhr{\textbf{SMOS $\uparrow$}} \\ 
\midrule
\multirow{2}{*}{AR+SoundStorm} 
& MLS & 
8.9\% & \textbf{0.612} & 49.11 &  -0.36 & 3.13 & %Libri
7.7\% & 0.587 & 20.76 &  0.09 & 3.71 \\ %Emi
& Emilia-En & 
\textbf{8.4\%} & 0.577 & \textbf{24.73} &  \textbf{-0.19} & \textbf{3.28} & %Libri
\textbf{6.6\%} & \textbf{0.618} & \textbf{12.73} & \textbf{0.19} & \textbf{3.73} \\ %Emi
\midrule
\multirow{2}{*}{VoiceBox} 
& MLS & 
\textbf{6.1\%} & \textbf{0.625} & \textbf{16.83} & 0.36 & 3.62 & %Libri
8.2\% & 0.528 & 15.94 & \textbf{0.28} & 3.61 \\ %Emi
& Emilia-En 
& 7.2\% & 0.585 & 23.24 & \textbf{0.42} & \textbf{3.77} &  %Libri
\textbf{7.4\%} & \textbf{0.601} & \textbf{14.07} & \textbf{0.28} & \textbf{3.76} \\ %Emi
\bottomrule
\end{tabular}
\end{table*}

\paragraph{Results and Discussions}

Table~\ref{tab:combined_evaluation} summarizes the objective and subjective evaluation results of TTS models trained on the Emilia-En and MLS datasets on the LibriSpeech-Test and Emilia-Test evaluation sets. 

On LibriSpeech-Test, the AR+SoundStorm model trained on Emilia-En achieved a lower WER (8.4\%) and FSD (24.73) compared to its MLS-trained counterpart, while the VoiceBox model trained on MLS achieved the best WER (6.1\%), S-SIM (0.625), and FSD (16.83). 
On the Emilia-Test, the AR+SoundStorm model trained on Emilia-En outperformed the MLS-trained model across all metrics, including WER (6.6\%), S-SIM (0.618), FSD (12.73), CMOS (0.19), and SMOS (3.73). Similarly, the Emilia-En trained VoiceBox achieved superior results in WER (7.4\%), S-SIM (0.601), and SMOS (3.76) compared to the MLS-trained version.

The results validate that, on the \hhr{formal reading style} LibriSpeech-Test evaluation set, models trained on both the Emilia-En and MLS datasets achieve comparable levels of intelligibility, speaker similarity, and naturalness. This suggests that Emilia, despite being derived from source speech data in-the-wild, is as effective as high-quality datasets like MLS after processing with our proposed Emilia-Pipe.
However, on the Emilia-Test evaluation set, which includes more spontaneous and human-like speech, training on in-the-wild datasets like Emilia significantly enhances the performance of speech generation models.

\paragraph{Summary (Answer to EQ1)}
The comparison between the in-the-wild Emilia dataset and audio-book MLS dataset for speech generation tasks reveals that while both types of datasets yield comparable performance in formal, audio-book-style speech, Emilia significantly outperforms audio-book datasets in generating more spontaneous and human-like speech, showcasing significantly superior performance in cloning diverse speaker timbre and speaking styles.

\subsection{Data Scaling Law in Speech Generation (EQ2)}
\hhr{To address EQ2}, we conduct experiments to investigate the impact of dataset size on speech generation performance with a fixed number of model parameters, i.e., the data scaling law in speech generation. 

\paragraph{Experimental Setups}
We leverage the \hhr{baseline models, evaluation sets, and objective metrics described in Sec.~\ref{sec:eq1}. To assess performance scaling, we progressively increase the training set size to the following amounts: 5k, 10k, 46k (the total duration of English speech in Emilia-En), 100k, and 134k hours (the total duration in Emilia-Large). At each stage, we record the corresponding changes in model performance.}

\paragraph{Results and Discussions}

\hhr{Fig.~\ref{fig:wer_scaling} shows the WER trends for AR+SoundStorm and VoiceBox as training set size increases. On LibriSpeech-Test (L), both models exhibit a decrease in WER; for instance, WER of AR+SoundStorm drops from 5.2\% at 5k hours to 4.2\% at 134k hours. Similarly, on the Emilia test set (E), the WER of AR+SoundStorm decreases from 5.7\% to 4.9\%, while the WER of VoiceBox declines from 7.1\% to 6.4\%.}

\hhr{Fig.~\ref{fig:sim_scaling} illustrates the S-SIM trends. On LibriSpeech-Test, AR+SoundStorm improves from 0.587 to 0.620, while VoiceBox rises from 0.418 to 0.606 as the dataset scales to 134k hours. On Emilia-Test, AR+SoundStorm increases from 0.587 to 0.636, and VoiceBox advances from 0.422 to 0.612.}

\hhr{Fig.~\ref{fig:fsd_scaling} shows the FSD trends. On LibriSpeech-Test, AR+SoundStorm decreases from 22.00 to 20.40, and VoiceBox decreases from 21.33 to 19.85. On Emilia-Test, AR+SoundStorm changes from 14.82 to 15.31, while VoiceBox decreases from 16.39 to 14.48.}

The results reveal a consistent scalability pattern across all metrics: both models demonstrate steady improvements as the dataset size increases, with only one exception of FSD for AR+SoundStorm (E). For smaller datasets (5k to 10k hours), performance gains are more pronounced, indicating that even modest increases in training data yield significant improvements. As the dataset size exceeds 46k hours, the rate of improvement slows but maintains a positive trend, eventually converging at approximately 100k hours.

\begin{figure*}[t!]
    \centering

    % Subfigure 1: WER
    \begin{subfigure}[t]{0.32\textwidth}
        \centering
        \resizebox{\linewidth}{!}{%
        \begin{tikzpicture}
            \definecolor{oiBlue}{HTML}{0072B2}
            \definecolor{oiOrange}{HTML}{E69F00}
            \definecolor{oiGreen}{HTML}{009E73}
            \definecolor{oiVerm}{HTML}{D55E00}
            \begin{axis}[
                ymin=3, ymax=9,
                symbolic x coords={5k, 10k, 46k, 100k, 134k},
                xtick=data,
                ymajorgrids=true,
                grid style=dashed,
                enlarge x limits=0.05,
                legend pos=north east,
                legend style={font=\footnotesize},
            ]
            \addplot[color=oiBlue, mark=o, line width=1.5pt, nodes near coords]
            coordinates {(5k, 5.2)(10k, 4.7)(46k, 4.4)(100k, 4.5)(134k, 4.2)};
            \addlegendentry{AR+SoundStorm (L)}

            \addplot[color=oiOrange, mark=o, line width=1.5pt, nodes near coords]
            coordinates {(5k, 7.2)(10k, 5.9)(46k, 5.7)(100k, 5.7)(134k, 5.6)};
            \addlegendentry{VoiceBox (L)}

            \addplot[color=oiGreen, mark=triangle, line width=1.5pt, nodes near coords]
            coordinates {(5k, 5.7)(10k, 5.2)(46k, 4.9)(100k, 4.9)(134k, 4.9)};
            \addlegendentry{AR+SoundStorm (E)}

            \addplot[color=oiVerm, mark=triangle, line width=1.5pt, nodes near coords]
            coordinates {(5k, 7.1)(10k, 6.5)(46k, 6.3)(100k, 6.7)(134k, 6.4)};
            \addlegendentry{VoiceBox (E)}
            \end{axis}
        \end{tikzpicture}%
        }
        \caption{WER (\%)}
        \label{fig:wer_scaling}
    \end{subfigure}
    \hfill
    % Subfigure 2: S-SIM
    \begin{subfigure}[t]{0.33\textwidth}
        \centering
        \resizebox{\linewidth}{!}{%
        \begin{tikzpicture}
            \definecolor{oiBlue}{HTML}{0072B2}
            \definecolor{oiOrange}{HTML}{E69F00}
            \definecolor{oiGreen}{HTML}{009E73}
            \definecolor{oiVerm}{HTML}{D55E00}
            \begin{axis}[
                ymin=0.4, ymax=0.8,
                symbolic x coords={5k, 10k, 46k, 100k, 134k},
                xtick=data,
                ymajorgrids=true,
                grid style=dashed,
                enlarge x limits=0.05,
                legend pos=north east,
                legend style={font=\footnotesize},
            ]
            \addplot[color=oiBlue, mark=o, line width=1.5pt, nodes near coords]
            coordinates {(5k, 0.587)(10k, 0.596)(46k, 0.613)(100k, 0.626)(134k, 0.620)};
            \addlegendentry{AR+SoundStorm (L)}

            \addplot[color=oiOrange, mark=o, line width=1.5pt, nodes near coords]
            coordinates {(5k, 0.418)(10k, 0.458)(46k, 0.596)(100k, 0.601)(134k, 0.606)};
            \addlegendentry{VoiceBox (L)}

            \addplot[color=oiGreen, mark=triangle, line width=1.5pt, nodes near coords]
            coordinates {(5k, 0.587)(10k, 0.589)(46k, 0.608)(100k, 0.635)(134k, 0.636)};
            \addlegendentry{AR+SoundStorm (E)}

            \addplot[color=oiVerm, mark=triangle, line width=1.5pt, nodes near coords]
            coordinates {(5k, 0.422)(10k, 0.443)(46k, 0.589)(100k, 0.581)(134k, 0.612)};
            \addlegendentry{VoiceBox (E)}
            \end{axis}
        \end{tikzpicture}%
        }
        \caption{S-SIM}
        \label{fig:sim_scaling}
    \end{subfigure}
    \hfill
    % Subfigure 3: FSD
    \begin{subfigure}[t]{0.33\textwidth}
        \centering
        \resizebox{\linewidth}{!}{%
        \begin{tikzpicture}
            \definecolor{oiBlue}{HTML}{0072B2}
            \definecolor{oiOrange}{HTML}{E69F00}
            \definecolor{oiGreen}{HTML}{009E73}
            \definecolor{oiVerm}{HTML}{D55E00}
            \begin{axis}[
                ymin=14, ymax=25,
                symbolic x coords={5k, 10k, 46k, 100k, 134k},
                xtick=data,
                ymajorgrids=true,
                grid style=dashed,
                enlarge x limits=0.05,
                legend pos=north east,
                legend style={font=\footnotesize},
            ]
            \addplot[color=oiBlue, mark=o, line width=1.5pt, nodes near coords]
            coordinates {(5k, 22.00)(10k, 21.97)(46k, 20.71)(100k, 20.58)(134k, 20.40)};
            \addlegendentry{AR+SoundStorm (L)}

            \addplot[color=oiOrange, mark=o, line width=1.5pt, nodes near coords]
            coordinates {(5k, 21.33)(10k, 21.23)(46k, 19.94)(100k, 19.93)(134k, 19.85)};
            \addlegendentry{VoiceBox (L)}

            \addplot[color=oiGreen, mark=triangle, line width=1.5pt, nodes near coords]
            coordinates {(5k, 14.82)(10k, 15.27)(46k, 15.62)(100k, 15.43)(134k, 15.31)};
            \addlegendentry{AR+SoundStorm (E)}

            \addplot[color=oiVerm, mark=triangle, line width=1.5pt, nodes near coords]
            coordinates {(5k, 16.39)(10k, 15.13)(46k, 14.65)(100k, 14.78)(134k, 14.48)};
            \addlegendentry{VoiceBox (E)}
            \end{axis}
        \end{tikzpicture}%
        }
        \caption{FSD}
        \label{fig:fsd_scaling}
    \end{subfigure}

    % Unified caption
    \caption{
        Model performance vs. training set size on LibriSpeech-Test (L) and Emilia-Test (E).
    }
    \label{fig:combined_scaling}
\end{figure*}
\paragraph{Summary (Answer to EQ2)}
\hhr{Our experimental results demonstrate a data scaling law in speech generation: as dataset size increases, performance improves, but with diminishing returns. Significant gains are observed when scaling up to 46k hours; beyond this point, improvements continue but become less substantial and tend to plateau around 100k hours. This insight can help guide future research in balancing dataset size and computational resources for optimal speech generation. For TTS models containing around 0.5–1 billion parameters,\footnote{The size of our baselines also falls within this range.} a dataset of approximately 100k hours per language appears to be the most cost-effective choice.}

\begin{table*}[ht]
\centering
\caption{Experimental results of AR+SoundStorm and VoiceBox for multilingual and crosslingual speech generation. The models were trained on the Emilia-Large dataset. Results for multilingual speech generation are highlighted in gray.}
\label{tab:multilingual}
\begin{tabular}{
  c  r  c c c c c c c c c c c c
}
\toprule
\multirow{2}{*}{
\diagbox{\textbf{Reference}}{\textbf{Target}}} & \multirow{2}{*}{\textbf{Metric}}& \multicolumn{6}{c}{\textbf{AR+SoundStorm}} & \multicolumn{6}{c}{\textbf{VoiceBox}} \\
\cmidrule(lr){3-8} \cmidrule(lr){9-14}
 & & {\textbf{En}} & {\textbf{Zh}} & {\textbf{Fr}} & {\textbf{De}} & {\textbf{Ja}} & {\textbf{Ko}} & {\textbf{En}} & {\textbf{Zh}} & {\textbf{Fr}} & {\textbf{De}} & {\textbf{Ja}} & {\textbf{Ko}} \\
\midrule
& \hhr{\textit{WER $\downarrow$}} & \cellcolor{lightgray!50}5.9\% & 5.8\% & 6.4\% & 5.9\% & 6.3\% & 8.3\% & \cellcolor{lightgray!50}6.5\% & 7.9\% & 8.8\% & 8.6\% & 8.3\% & 10.2\% \\
\textbf{En} & \hhr{\textit{S-SIM $\uparrow$}} & \cellcolor{lightgray!50}0.568 & 0.431 & 0.452 & 0.529 & 0.446 & 0.443 & \cellcolor{lightgray!50}0.588 & 0.386 & 0.458 & 0.490 & 0.425 & 0.442 \\
& \hhr{\textit{FSD $\downarrow$}} & \cellcolor{lightgray!50}24.99 & 99.40 & 82.84 & 26.62 & 89.40 & 98.36 & \cellcolor{lightgray!50}24.34 & 91.29 & 78.53 & 68.62 & 92.54 & 89.49 \\
\midrule
& \hhr{\textit{WER $\downarrow$}} & 5.3\% & \cellcolor{lightgray!50}3.6\% & 5.2\% & 5.4\% & 4.9\% & 5.7\% & 8.6\% & \cellcolor{lightgray!50}5.6\% & 6.7\% & 5.9\% & 6.4\% & 7.0\% \\
\textbf{Zh} & \hhr{\textit{S-SIM $\uparrow$}} & 0.507 & \cellcolor{lightgray!50}0.511 & 0.509 & 0.504 & 0.516 & 0.523 & 0.524 & \cellcolor{lightgray!50}0.557 & 0.524 & 0.522 & 0.543 & 0.591 \\
& \hhr{\textit{FSD $\downarrow$}} & 56.15 & \cellcolor{lightgray!50}40.09 & 56.75 & 57.10 & 56.71 & 52.60 & 109.67 & \cellcolor{lightgray!50}40.04 & 58.47 & 72.47 & 64.73 & 57.90 \\
\midrule
& \hhr{\textit{WER $\downarrow$}} & 5.3\% & 5.3\% & \cellcolor{lightgray!50}5.3\% & 5.2\% & 5.8\% & 8.1\% & 7.0\% & 6.3\% & \cellcolor{lightgray!50}5.6\% & 6.9\% & 7.5\% & 9.3\% \\
\textbf{Fr} & \hhr{\textit{S-SIM $\uparrow$}} & 0.596 & 0.527 & \cellcolor{lightgray!50}0.596 & 0.596 & 0.572 & 0.557 & 0.565 & 0.485 & \cellcolor{lightgray!50}0.589 & 0.582 & 0.547 & 0.556 \\
& \hhr{\textit{FSD $\downarrow$}} & 39.89 & 66.21 & \cellcolor{lightgray!50}39.88 & 38.48 & 51.13 & 54.41 & 91.08 & 80.76 & \cellcolor{lightgray!50}42.38 & 58.16 & 63.36 & 61.51 \\
\midrule
& \hhr{\textit{WER $\downarrow$}} & 4.5\% & 4.5\% & 4.7\% & \cellcolor{lightgray!50}4.2\% & 4.8\% & 6.8\% & 5.2\% & 7.4\% & 6.8\% & \cellcolor{lightgray!50}5.2\% & 6.9\% & 8.9\% \\
\textbf{De} & \hhr{\textit{S-SIM $\uparrow$}} & 0.619 & 0.545 & 0.603 & \cellcolor{lightgray!50}0.639 & 0.596 & 0.591 & 0.639 & 0.519 & 0.577 & \cellcolor{lightgray!50}0.683 & 0.538 & 0.586 \\
& \hhr{\textit{FSD $\downarrow$}} & 39.96 & 57.82 & 44.86 & \cellcolor{lightgray!50}33.16 & 53.38 & 55.12 & 83.37 & 72.18 & 54.77 & \cellcolor{lightgray!50}34.41 & 67.89 & 67.46 \\
\midrule
& \hhr{\textit{WER $\downarrow$}} & 4.6\% & 4.4\% & 4.7\% & 4.5\% & \cellcolor{lightgray!50}4.8\% & 6.6\% & 7.4\% & 5.5\% & 6.9\% & 6.7\% & \cellcolor{lightgray!50}6.2\% & 6.6\% \\
\textbf{Ja} & \hhr{\textit{S-SIM $\uparrow$}} & 0.622 & 0.557 & 0.626 & 0.618 & \cellcolor{lightgray!50}0.641 & 0.633 & 0.556 & 0.525 & 0.521 & 0.557 & \cellcolor{lightgray!50}0.584 & 0.596 \\
& \hhr{\textit{FSD $\downarrow$}} & 49.42 & 68.70 & 44.67 & 50.47 & \cellcolor{lightgray!50}44.28 & 52.19 & 103.68 & 76.65 & 63.55 & 72.41 & \cellcolor{lightgray!50}44.71 & 56.34 \\
\midrule
& \hhr{\textit{WER $\downarrow$}} & 6.2\% & 4.1\% & 6.1\% & 6.2\% & 6.2\% & \cellcolor{lightgray!50}6.3\% & 8.0\% & 5.6\% & 7.8\% & 8.3\% & 5.6\% & \cellcolor{lightgray!50}6.0\% \\
\textbf{Ko} & \hhr{\textit{S-SIM $\uparrow$}} & 0.657 & 0.593 & 0.665 & 0.656 & 0.673 & \cellcolor{lightgray!50}0.673 & 0.589 & 0.567 & 0.545 & 0.597 & 0.595 & \cellcolor{lightgray!50}0.648 \\
& \hhr{\textit{FSD $\downarrow$}} & 36.71 & 58.85 & 32.27 & 37.20 & 31.95 & \cellcolor{lightgray!50}30.27 & 86.57 & 63.49 & 53.75 & 57.19 & 52.85 & \cellcolor{lightgray!50}38.82 \\
\bottomrule
\end{tabular}
\end{table*}

\subsection{Multilingual and Crosslingual Speech Generation (EQ3)}
\hhr{To address EQ3, we conduct experiments to evaluate the effectiveness of Emilia for training both multilingual (where the reference and target speech are in the same language) and crosslingual (where the reference and target speech are in different languages) speech generation models.}

\paragraph{Experimental Setup} 
We utilize the complete Emilia-Large dataset, which encompasses six languages: English (En), Chinese (Zh), German (De), French (Fr), Japanese (Ja), and Korean (Ko), to train speech generation models. 
For English evaluation, we leverage Emilia-Test. The Chinese evaluation set is randomly sampled from the AISHELL-3 dataset \cite{aishell3}. The evaluation sets for German, French, Japanese, and Korean are sourced from Common Voice \cite{common_voice}. 
\hhr{Each evaluation set contains at least 500 reference speech samples. In multilingual experiments, these reference speech samples are used to synthesize target texts in the same language. For crosslingual experiments, the reference speech samples and target texts are selected from evaluation sets of different languages.}

\paragraph{Results and Discussions}

The experimental results in Table~\ref{tab:multilingual} demonstrate the effectiveness of the Emilia-Large dataset for multilingual and crosslingual speech generation. 

In multilingual generation, where the reference and target languages are the same, both AR+SoundStorm and VoiceBox achieve strong performance across all six languages. \hhr{For example, AR+SoundStorm attains WERs ranging from 3.6\% (Zh-Zh) to 6.3\% (Ko-Ko), with S-SIM scores between 0.511 (Zh-Zh) and 0.673 (Ko-Ko), and FSD as low as 24.99 (En-En).}

In crosslingual generation, where the reference and target languages differ, model performance shows moderate degradation. \hhr{For instance, AR+SoundStorm's WER increases from 6.3\% (Ko-Ko) to 8.3\% (En-Ko). Similarly, VoiceBox's S-SIM decreases from 0.588 (En-En) to 0.386 (En-Zh). Nonetheless, the overall performance of both models in crosslingual generation remains competitive.}
These results validate the effectiveness of the Emilia-Large dataset in training strong multilingual and crosslingual speech generation models. 

% Second, crosslingual generation introduces notable challenges: while models retain functionality, performance gaps emerge, particularly for VoiceBox, indicating that the NAR-based architectures may struggle more with cross-language acoustic transfer.
%
Furthermore, when comparing model performance to those trained exclusively on English data, the multilingual models exhibit a slight trade-off in English generation. \hhr{For example, AR+SoundStorm trained on the 216k-hour multilingual Emilia-Large dataset achieves a WER of 4.9\%, an S-SIM of 0.636, and an FSD of 15.31 on English evaluation samples. These metrics are marginally worse than those of its monolingual counterpart trained on a 134k-hour English-only dataset (WER=4.5\%, S-SIM=0.65, FSD=14.8, as reported in Fig.~\ref{fig:wer_scaling}--\ref{fig:fsd_scaling}).} This suggests that while multilingual data enables crosslingual capabilities, it can compromise language-specific performance. Such trade-offs are important considerations for practical applications, and future work could explore strategies to narrow these performance gaps.

\paragraph{Summary (Answer to EQ3)}
\hhr{The Emilia-Large dataset can be effectively used to train robust multilingual and crosslingual speech generation models.} However, crosslingual generation introduces moderate performance degradation, highlighting challenges in cross-language acoustic transfer.
These findings underscore the value of the Emilia-Large dataset as a critical resource for advancing multilingual and crosslingual speech generation. Future work could focus on enhancing model adaptability to address crosslingual challenges.

\section{Conclusion and \hhr{Discussion}}\label{sec:conclusion}
In conclusion, this work first introduces Emilia-Pipe, an effective and efficient preprocessing pipeline designed to transform source speech data in-the-wild into high-quality training datasets for spontaneous and human-like speech generation. 
Leveraging Emilia-Pipe, we construct Emilia, \hhr{one of the largest open-source multilingual speech generation datasets}, spanning over 101k hours across six languages, as well as its extended version, Emilia-Large, which contains 216k hours of data.
Comparative analyses demonstrate that Emilia significantly outperforms traditional audio-book datasets in generating spontaneous and human-like speech. Our experiments also investigate the relationship between dataset size and speech generation performance, revealing consistent improvements with data scaling, though the trend becomes less pronounced as the dataset size exceeds 100k hours. Finally, we validate that the proposed Emilia dataset effectively supports multilingual and crosslingual speech generation, paving the way for future advancements in this field. 

Future work may focus on training effective spoof detection models to address potential safety concerns associated with highly spontaneous and human-like speech generation models trained on Emilia, such as the risk of synthetic spoken misinformation \cite{spmis,overtheair}. Additionally, expanding Emilia to include the singing/music domain could benefit singing/music voice generation \cite{singnet}.

Despite the advancements, we point out a few limitations. First, the speaker diarization model we use is not perfect and can result in a small proportion of speech segments containing more than one speaker or overlaps. This issue can subsequently affect speech generation performance. Integrating stronger models in the future may alleviate this issue.
Second, Emilia-Pipe segments speech samples into intervals of 3 to 30 seconds. It is observed that generating speech outside this range may lead to unexpected outcomes. Adjustments in the hyper-parameters of Emilia-Pipe may be needed for specific use cases.
\hhr{Third, due to our resource constraints, the current Emilia-Large dataset covers only six languages. We welcome community contributions to extend the dataset to more languages with our open-source Emilia-Pipe to enhance its global applicability.}

\bibliographystyle{IEEEtran}
\bibliography{main}

\end{document}